# Democratizing Design for Future Computing Platforms

Luis Ceze, Mark D. Hill, Karthikeyan Sankaralingam, and Thomas F. Wenisch

**Abstract**
Information and communications technology can continue to change our world. These advances will partially depend upon designs that synergistically combine software with specialized hardware. Today open-source software incubates rapid software-only innovation. The government can unleash software-hardware innovation with programs to develop open hardware components, tools, and design flows that simplify and reduce the cost of hardware design. Such programs will speed development for startup companies, established industry leaders, education, scientific research, and for government intelligence and defense platforms.

Information and communications technology (ICT) is changing our world. And yet, even more transformative advances are possible, for example, by using data to personalize medicine and education to the needs of individuals. With advances in machine learning and cloud computing, future ICT improvements will likely exceed past advances, bringing improved quality of life and enhanced competitiveness for the USA as world leader.

Past ICT gains have been facilitated by synergistic improvements in hardware, commonly called "Moore's Law." Moore's Law enabled ICT hardware performance and cost-performance to double every two years to provide a foundation for the ICT wonders we all experience. Unfortunately, the technology trends fueling Moore's Law have run against critical physical constraints. Today, this foundation of ICT innovation is no longer rapidly improving, especially in energy-constrained environments from smartphones to data centers.

Another key ICT foundation has been the explosion of software innovations incubated by the open-source movement. Open-source software allows universities, established companies, and startups to begin with an improved software infrastructure—such as the LAMP stack (Linux, Apache, MySQL, & PHP) for providing services over the web. Open source software is especially important for start-up companies because it enables them to launch with much software in place and then rapidly add their "special sauce" to succeed (e.g., Facebook) or fail fast in months rather than requiring multiple rounds of investment. Open source software reduces risk and accelerates the impact of capital investment.

To overcome the reduced improvement in conventional hardware, many innovators are looking at designing hardware and software together rather than separately, as has been the norm the last several decades. Such cross-layer design can enable specialization that targets specific use cases with effective, energy-efficient designs that are especially important in emerging low-energy opportunities, such as medical monitoring using energy harvesting (e.g., from motion). In some cases, the line between software and hardware will continue to blur with software-configurable hardware fabrics like field-programmable gates arrays (FPGAs).

Without hardware-software co-design, software-only designs will be relegated to less effectiveness due a "specialization gap." For example, without specialization, it may be beyond our reach to do continuous mobile computer vision, real time virtual reality video for telepresence, and sequencing and analyzing the genomes of everyone one (or every living thing) on the planet. These barriers are there because current design--with largely separate software and hardware--is insufficient and technology improvement to hardware has slowed dramatically.

Unfortunately, hardware-software co-designers must currently eschew many of the benefits of open-source, because the hardware aspects of systems are decidedly propriety. For example, a hardware-software startup for a wearable Internet-of-Things system must spend considerable time and money licensing even the most basic hardware components and computer-aided design tools.  As a result, bringing a hardware-software product to market requires as much as ten times the venture investment as compared to, e.g., a software-only phone application.  The steep investment requirement severely limits the target markets for hardware-software innovation.

Wouldn't it be better if future hardware-software advances could be made with the rapidity that heretofore was limited to software only?
1. This will allow the private sector to experiment in providing value quickly, but sometime failing fast and inexpensively.
2. It will make viable investment in products with smaller natural markets, such as a medical devices for less common ailments or assistive devices for the hearing/sight impaired.
3. It will enable researchers—from science to medicine—to build better systems to monitor and act, e.g., just enough insulin.
4. It will facilitate education by enabling students at many levels to work with open tools and designs.
5. It can help the public sector by allowing design from a high baseline for applications in security, intelligence and defense that are rapidly changing and not mass market. Today, such systems are frequently plagued with serious schedule and cost overruns.

Without government action, Americans may miss the opportunity to benefit from innovative hardware-software devices due to barriers caused by lack of reusability of designs and poor ease of use of tools. Moreover, companies with proprietary building blocks and proprietary computer-aided design tools have local incentives to resist democratizing hardware-software design, much as established software companies resisted open-source software decades ago.

For these reasons, we recommend that government promote continued hardware-software innovation to "democratize" hardware with shared open-source building blocks and computer-aided design tools that can enable many—even students and tinkerers— to contribute to hardware innovation in an open and inspectable manner. The government can catalyze this effort with programs to develop hardware components, tools, and design flow. Initial work might focus on less-expensive, stable, older technologies suitable to innovative hardware-software systems in the Internet of Things. The task also includes challenges beyond software-only systems, especially because hardware eventually has to be manufactured.

While the exact role of government, academic, and industrial stakeholders will have to be determined, here are some compelling possibilities. First, the government can fund or otherwise encourage the development of hardware components with open intellectual property (IP) rights, perhaps through explicit grants or encouragement for "broader impacts". There are already several initial forays along these lines. On one hand, RISC V [https://riscv.org/] provides an open instruction set architecture--with several variations--they can be publically implemented and extended. On the other hand, DARPA CHIPS [DARPA-BAA-16-62] focuses on developing reusable IP--including physically reusable chiplets--that are commercially available, but not necessarily open.

Second, the government can also encourage open computer-aided design (CAD) tools. Currently one of the biggest impediments to the success of open-source hardware is that design tools are expensive, typically proprietary, and quite obfuscated in explanation of how to be used. While some open-source tools like Verilator (for behavioral simulation) and Yosys (for synthesis) exist, design tools are nowhere near the level of software engineering maturity of Linux, glibc and make for software. Well packaged, and license-free open-source development tools that span RTL, synthesis, backend design, and some forms of backend IP will be necessary and can make a transformative impact to open-source hardware. As an analogy, when linux and gcc became mature and competitive with proprietary operating systems and compilers, they very quickly surpassed proprietary tools in capability and usage, leading to today's cloud, data-center, and billion-dollar startups like Google and Facebook. We needs efforts to develop open-source, well-packaged, modular tools for hardware design.

Third, the government can encourage and support more chip building--creating funding programs that allow research teams to build chips will help drive open-source hardware efforts and build compelling prototypes. Building usable physical hardware artifacts will make open-source hardware compelling for developers to contribute and for industry practitioners to leverage. In that regard, enabling and simplifying the path to build hardware can be transformative. Due to the slowing down of Moore's Law (more transistors per chip) and Dennard's scaling (each transistor better), benefits of newer technology nodes have been small. On the positive side, the tremendous volumes at the 28nm technology node make it very affordable to build chips (as research prototype and for startups). Companies like eSilicon serve as intermediaries to foundries for building chips at 28nm for as "little" as $145,000 for small chips in a multi-project wafer. Alternatively, MOSIS [https://www.mosis.com/] could be expanded with an emphasis on open hardware IP.

Modest government action can seed a virtuous cycle. Arthur C. Clarke said, "Any sufficiently advanced technology is indistinguishable from magic." Let's make some more magic!

*This material is based upon work supported by the National Science Foundation under Grant No. 1136993. Any opinions, findings, and conclusions or recommendations expressed in this material are those of the authors and do not necessarily reflect the views of the National Science Foundation.*

*For citation use: Ceze L., Hill M., Sankaralingam K., & Wenisch T. (2017) Democratizing Design for Future Computing Platforms. http://cra.org/ccc/resources/ccc-led-whitepapers/*